\documentclass[prl,aps,epsf,showpacs,twocolumn]{revtex4}

\usepackage{times}
\usepackage{graphicx}
\usepackage{float}
\usepackage{latexsym,amsmath,amssymb,bm,euscript}
\usepackage{color}
\usepackage{subfigure}
\usepackage{epstopdf}
\usepackage[colorlinks=true,linkcolor=blue,citecolor=blue]{hyperref}
\usepackage{hyperref}
\usepackage{ulem}

\usepackage{appendix}

\begin{document}

\title{Chiral and Critical Spin Liquids  in Spin-$1/2$ Kagome Antiferromagnet}
\author{W. Zhu, S. S. Gong, and D. N. Sheng}
\affiliation{Department of Physics and Astronomy, California State University, Northridge, California 91330, USA}

\begin{abstract}
The topological quantum spin liquids (SL) and the nature of quantum phase transitions between them  have attracted
intensive attentions for the past twenty years. The extended kagome spin-1/2 antiferromagnet emerges as the
primary candidate for hosting both time reversal symmetry (TRS) preserving and TRS  breaking SLs
based on density matrix renormalization group simulations.
To uncover the nature of the novel quantum phase transition between the  SL states,
we study a minimum  XY model  with the nearest neighbor (NN) ($J_{xy}$), the  second and third
NN couplings ($J_{2xy}=J_{3xy}=J'_{xy}$).
We identify the TRS broken  chiral SL (CSL) with the turn on of a small perturbation $J'_{xy}\sim 0.06 J_{xy}$,
which is fully characterized by the fractionally  quantized topological Chern number and the
conformal edge spectrum as the $\nu=1/2$
fractional quantum Hall state. On the other hand, the NN XY model ($J'_{xy}=0$) is shown to be
a critical SL state adjacent to the CSL,  characterized by the gapless spin singlet excitations
and also vanishing small  spin triplet excitations.
The quantum phase transition from the CSL to the gapless critical SL is driven by
the collapsing of the neutral (spin singlet) excitation gap.
By following the evolution of entanglement spectrum,
we find that the  transition takes place through
the coupling  of  the edge states with opposite chiralities, which merge into the bulk
and become gapless neutral excitations.
The  effect of the NN spin-$z$ coupling $J_z$ is also studied,
which leads to a quantum phase diagram  with an extended regime for the gapless SL.
\end{abstract}

\maketitle


Quantum spin liquid (SL) is an exotic state of matter which escapes from forming the conventional orders even at zero temperature \cite{Nature_464_199}.
However, different from a featureless insulator, a SL develops a topological order \cite{PRB_40_7387, PRB_41_9377, IJMPB_4_239}
with fractionalized quasiparticles encoded in the long-range entanglement of system \cite{PRB_82_155138}.
The  SL physics may play a  fundamental role for understanding 
strongly correlated systems  and unconventional
superconductivity \cite{Science_235_1196, PRL_61_2376, PRL_66_1773, PRB_44_2664, PRB_62_7850, PRL_86_1881, PRB_65_224412, PRB_66_205104, PRL_89_277004,
PRL_94_146805, RMP_78_17, AP_321_2, PRL_99_097202, Science_321_1306, NP_7_772, PRL_108_247206}.
There have been intensive studies searching for  SL in frustrated magnetic systems,
however, the discovery of SL has been rare in the past 20 years.
A few frustrated square or  honeycomb lattices spin systems with competing interactions
have been proposed as the candidates for gapped SL
\cite{PRB_82_024419, PRB_84_024420, PRL_107_087204, PRB_86_024424}. However, further studies find
that the competing plaquette valence-bond solid may dominate the magnetic disorder region \cite{PRL_110_127203, PRL_110_127205,
PRB_88_165138, PRL_113_027201}.

Interestingly, the nearest neighbor (NN) dominant  spin-$1/2$ kagome Heisenberg model 
has been identified to host a gapped SL
based on the state of art density matrix renormalization group (DMRG) calculations \cite{PRL_101_117203, Science_332_1173, PRL_109_067201, NP_8_902},
where a near quantized topological entanglement entropy \cite{PRL_96_110404, PRL_96_110405} has been found consistent with a $Z_2$ SL \cite{PRL_109_067201,NP_8_902}.
The topological degeneracy as a signature evidence for such a gapped topological state
\cite{PRL_66_1773,PRB_44_2664, PRB_40_7387, PRB_60_1654,PRB_62_7850}
has not been established,  while different methods have been applied to tackle this
problem \cite{Science_332_1173,PRB_89_075110}.
Meanwhile, the variational studies find that the Dirac gapless SL
has the lower variational energy among different states
based  on  the projected fermionic parton wavefunctions \cite{PRL_98_117205, PRB_84_020407}.
The nature of the SL in the kagome Heisenberg model remains not fully understood.

By introducing the second- and third- NN couplings for the spin-$1/2$ kagome systems,
DMRG studies \cite{PRL_112_137202, SR_4_6317} discover the Kalmeyer-Laughlin CSL
theoretically predicted  more than 20 years ago \cite{PRL_50_1395, PRL_59_2095, PRB_39_11413, PRL_70_2641, PRB_52_4223}, which spontaneously breaks
TRS and is identified as the $\nu = 1/2$ fractional quantum Hall (FQH) state \cite{PRL_50_1395, PRL_59_2095, 1407.0869}.
Interestingly, the  CSL state is also found in the spin anisotropic kagome model involving the second and third NN
$xy$-plane couplings \cite{1407.2740}, or by introducing the
TRS breaking three-spin chiralities interactions \cite{1401.3017}.
However,  the nature of the quantum phase transition, especially how the
quantum state and entanglement spectrum (ES) evolve near such  a transition  have not been addressed.
We  do not know what a physical mechanism can drive the quantum phase transition in such a system,
and  if the emergence of the previously identified gapped SL for the NN kagome Heisenberg model has close
connection with the collapsing of the CSL \cite{1307.8194}.
Our work is motivated to address these open questions.

Along with theoretical developments, experiments also discover different promising SL candidates in the triangular organic
compounds \cite{PRL_91_2003, PRL_95_177001, PRB_77_2008} and kagome antiferromagnets Herbertsmithite and Kapellasite
Cu$_3$Zn(OH)$_6$Cl$_2$ in recent years \cite{PRL_98_077204, PRL_98_107204, PRL_103_237201, PRL_109_037208, Nature_492_7429, PRL_110_207208}.
These materials appear to have gapless excitations as revealed by the
specific heat and neutron scattering measurements \cite{PRL_98_107204, PRL_103_237201, PRL_109_037208, Nature_492_7429}.
Thus, it would be extremely interesting to also search for some  minimum spin-$1/2$ kagome model which can host
a gapless SL.

\begin{figure}
\includegraphics[width=0.4\linewidth]{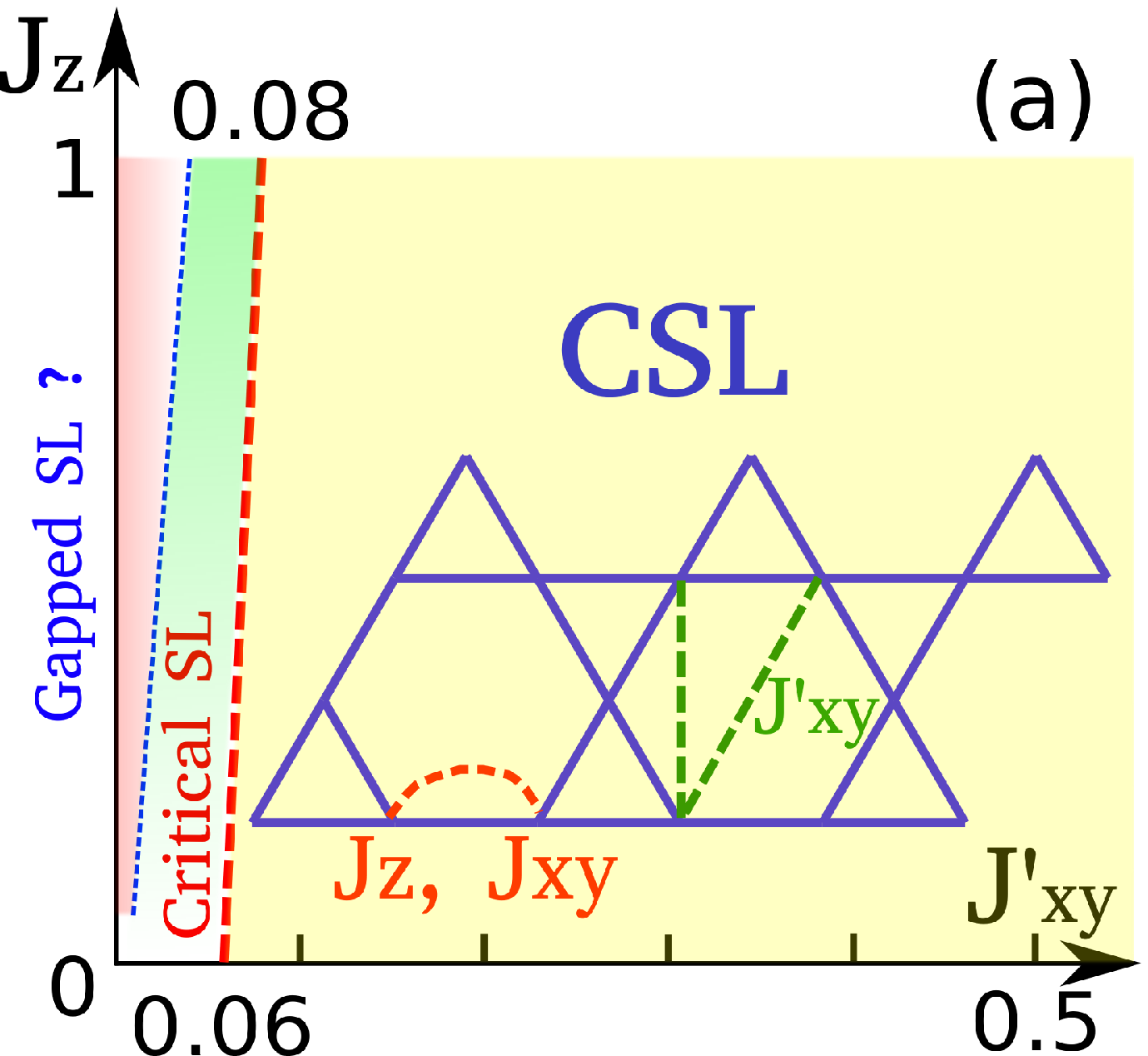}
\includegraphics[width=0.43\linewidth]{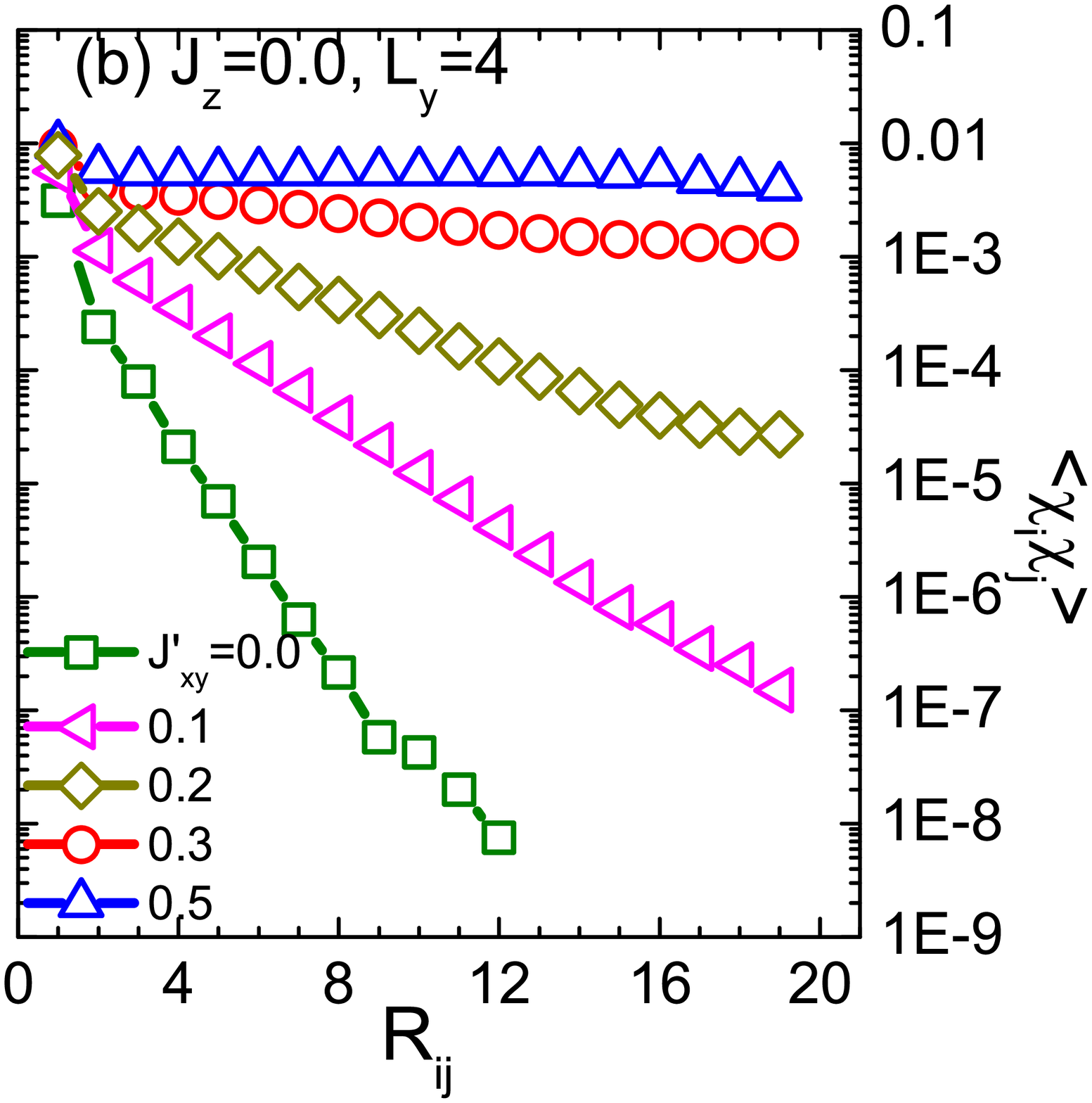}
\includegraphics[width=0.8\linewidth]{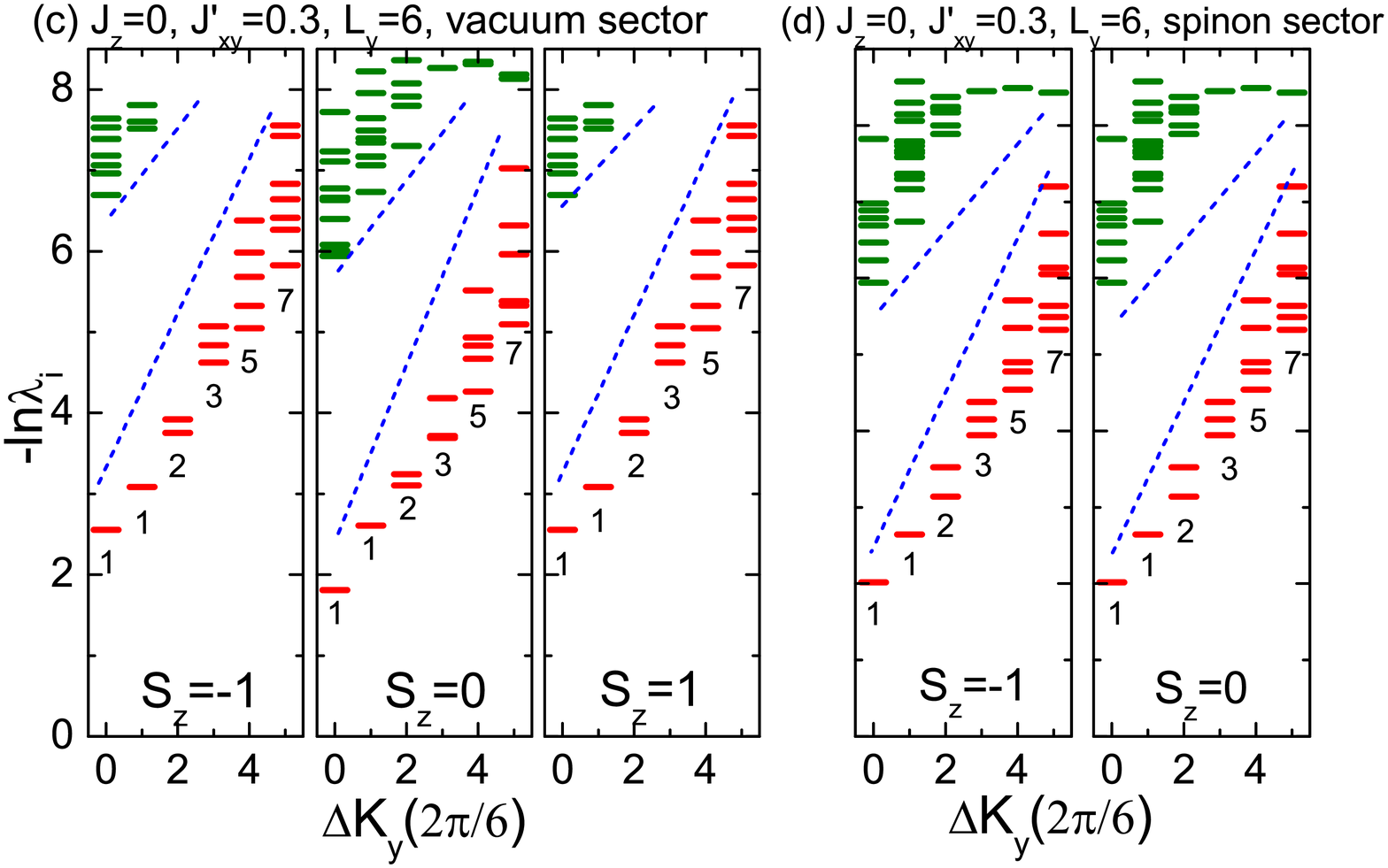}
\caption{(color online) (a) Schematic phase diagram of the spin-$1/2$ kagome model with Hamiltonian Eq. (\ref{Hamiltonian}).
for $0.0 \leq J_{xy}^{\prime} \leq 0.5$, $0.0 \leq J_z \leq 1.0$,
where both the CSL and a critical SL with gapless singlet excitations are identified.
The color gradient in the critical SL
denotes the growing spin gap on finite-size system with increasing $J_z$.
The gapped SL \cite{Science_332_1173} may exist neighboring with the critical SL.
(b) Log-linear plot of chiral correlation $\langle \chi_i \chi_j \rangle$ versus the
distance of triangles $R_{ij}$ along the $x$ direction in the vacuum sector 
(c) and (d) are the ES of the groundstates in the vacuum and spinon sectors, respectively.
$\lambda_i$ is the eigenvalue of reduced density matrix.
The numbers $\{1,1,2,3,5,7,\cdots\}$ label the near degenerating pattern
for the low-lying ES with different
relative momentum $\Delta k_y$ and total spin $S^z$. 
} \label{Fig1}
\end{figure}

In this Letter, we  address the nature of the collapsing of CSL and the related phase transition in kagome spin system
based on DMRG and exact diagonalization (ED) calculations.
We study the spin-$1/2$ XXZ kagome model with the spin XY interactions for the second and third neighbors
as shown in the inset of Fig.~\ref{Fig1}(a),
whose Hamiltonian is given as \cite{1407.2740}
\begin{eqnarray}
H &=& (J_{xy}/2) \sum_{\langle i,j\rangle} (S_{i}^{+} S_{j}^{-} + h.c.) + J_z \sum_{\langle i,j\rangle} S_{i}^{z} S_{j}^{z} \nonumber \\
&+& (J_{xy}^{\prime}/2) \sum_{\langle i,j\rangle ^{\prime}} (S_{i}^{+} S_{j}^{-} + h.c.), \label{Hamiltonian}
\end{eqnarray}
where the summations are taken over the NN $\langle i,j\rangle$, the second and third NN $\langle i,j\rangle ^{\prime}$ couplings.
We set $J_{xy}=1$ as energy scale. Our main results are summarized as the phase diagram in Fig.~\ref{Fig1}(a).
First of all, for the XY model with $J_z = 0$,
we establish a CSL for $J_{xy}^{\prime} \gtrsim 0.06$ based on the
topological features of the state:
the conformal chiral edge spectrum in accordance with the $\nu=1/2$ FQH state and
the topological quantized Chern number $C=1/2$.
We identify the physical driving force for the destruction  of the CSL as the collapsing of the singlet
excitation gap with reducing $J'_{xy}$.
Following the evolution of ES,
we find that the phase transition takes place through the coupling between the low-lying entanglement states
with opposite chiralities, which naturally leads to a critical state with TRS and gapless
neutral excitations \cite{PRL_84_3950, PRB_53_15845}.
Our results represent a significant progress in understanding the connection between different SLs \cite{1407.2740}
by identifying the mechanism of the phase transition and
establishing the characteristic nature of the critical SL phase adjacent to the CSL.
After tuning on the NN spin-$z$ coupling $J_z$, we identify the phase diagram for different $J'_{xy}$,
where the critical gapless SL is found for an extended regime with small $J'_{xy}\sim 0$.
The connection of the critical SL with the previously identified gapped SL in Heisenberg model \cite{Science_332_1173}
will also be discussed.


We use DMRG \cite{PRL_69_2863} and ED
to study cylinder and torus systems with the geometry shown in the inset of Fig.~\ref{Fig1}(a).
The number of sites in cylinder (open boundary condition in the $x$-direction) or torus system
is  $N = 3\times L_y \times L_x$
with $L_x$ and $L_y$ as the numbers of unit cell in the $x$ and $y$ directions \cite{suppl}.
We perform the flux insertion simulations on cylinder systems based on the newly developed adiabatic DMRG to detect
the topological Chern number \cite{SR_4_6317}.
In this simulation, we thread a flux $\theta$ in the cylinder
which is equivalent to imposing the twist boundary conditions:
$S_i^+S_j^-+h.c.  \rightarrow e^{i\theta_{ij}} S_i^+S_j^-+h.c.$  for these bonds crossing
the $y$ boundary.
In DMRG studies we keep up to $8000-10000$ states for the simulations without flux and $4500-6000$ states
in the flux insertion simulations for accurate results.

\textit{Chiral spin liquid phase.---}
The CSL breaks TRS but preserves lattice symmetries and spin rotational symmetry.
The TRS broken is usually detected by the chiral order parameter 
 $ \chi_i  =  (S_{i_1} \times S_{i_2})\cdot S_{i_3} $
($i_1,i_2,i_3 \in \bigtriangleup_i (\bigtriangledown_i)$ triangle) \cite{PRB_39_11413}.
As shown in Fig.~\ref{Fig1}(b), we demonstrate the chiral correlations
$\langle \chi_{i} \chi_{j} \rangle$ between the up-triangles $i$ and $j$ as a function of distance $R_{ij}$
for the XY model ($J_z=0$).
At $J_{xy}^{\prime} = 0$, the chiral correlations decay exponentially to vanish.
With growing $J_{xy}^{\prime}$,
the chiral correlations enhance gradually and appear to approach finite values for $J_{xy}^{\prime} \gtrsim 0.2$ at large distance,
which indicates the emerging long-range chiral order that characterizes the spontaneous TRS breaking.

Moreover, CSL is a topological ordered state that hosts two-fold topological degenerate groundstates,
which can be obtained by inserting flux with $\theta = 0  \rightarrow 2\pi$ adibatically \cite{SR_4_6317}.
The ES for these two states, labeled by the quantum number total $S^z$ of the half system,
and their relative momentum quantum number along the $y$ direction $\Delta k_y$
\cite{Vidal, PRL_110_236801}, are shown in Figs.~\ref{Fig1}(c) and \ref{Fig1}(d).
The leading ES has the robust degeneracy pattern $\{1,1,2,3,5,7,\cdots\}$
with increasing $\Delta k_y$ in each $S^z$ sector, which follows
the chiral $SU(2)_1$ Wess-Zumino-Witten conformal field theory description of the $\nu = 1/2$ FQH state \cite{CFT_book}
as the fingerprint for the emergence of the CSL \cite{PRL_101_010504}.
The spectra of the vacuum ($\theta = 0$) and spinon ($\theta=2\pi$) sectors are symmetric
about $S^z=0$ and $-\frac{1}{2}$ respectively, indicating a spin-$\frac{1}{2}$ spinon at each end of cylinder in the spinon sector.

\begin{figure}
\includegraphics[width=0.8\linewidth]{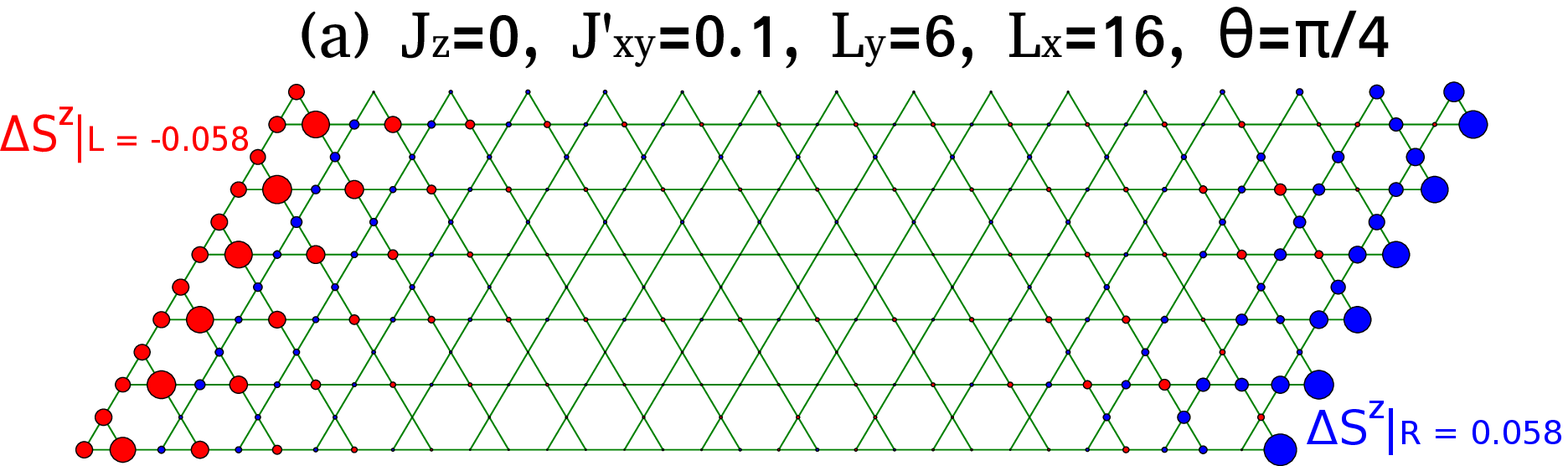}
\vspace{5pt}\\
\includegraphics[width=0.9\linewidth]{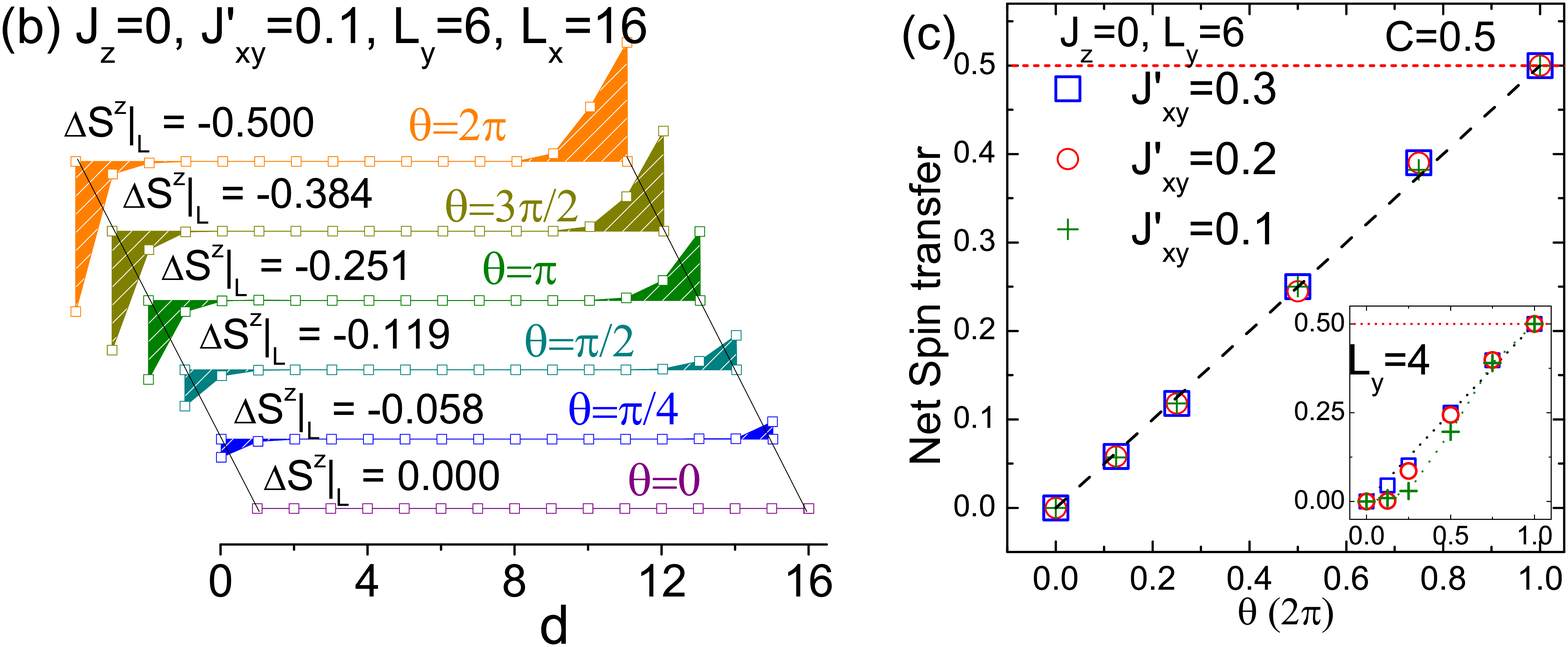}
\includegraphics[width=0.5\linewidth]{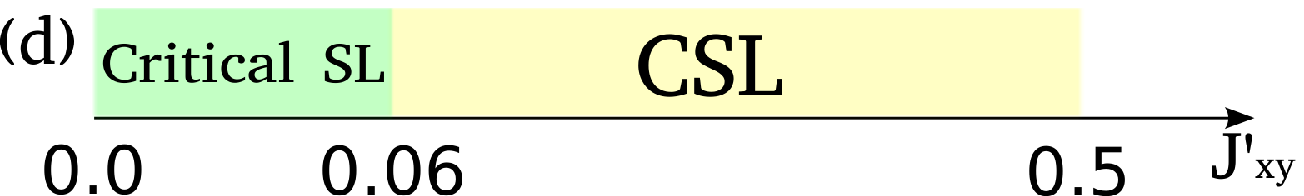}
\caption{(color online) (a) Spin magnetization $\langle S^z_{x,y}\rangle$ at $R_i=(x,y)$
after adiabatically inserting a flux $\theta=\pi/4$. The area of the circle is proportional to the amplitude
of $\langle S^z_{x,y}\rangle$. The blue (red) color represents the positive (negative) $\langle S^z_{x,y}\rangle$.
(b) Accumulated spin magnetization $\langle S^z_x\rangle = \sum_{y}\langle S^z_{x,y}\rangle$ in each column
(the summation is over all the $3L_y$ sites for each column) with increasing flux $\theta$.
(c) Net spin transfer $\Delta S^z|_{\rm edge}$ as a function of $\theta$ on $L_y=6$ cylinder.
The inset shows the results on $L_y=4$ cylinder.
(d) Phase diagram of the XY kagome model ($J_z = 0$),
which is determined from the results of Chern number on $L_y=6$ system.
} \label{chern}
\end{figure}


\textit{Fractional quantization of topological Chern number.---}
To reveal the full topological nature of the CSL phase, we perform the
flux insertion simulation to obtain the topological Chern number \cite{SR_4_6317}.
By adiabatically inserting flux $\theta$,
we study the evolution of the spin-$z$ magnetization $\langle S^z_{x,y}\rangle$ at each site $R_i=(x,y)$.
One example with a  small
$\theta = \pi/4$ is shown in Fig.~\ref{chern}(a).
We find that the nonzero magnetizations start to build up at both edges of cylinder after inserting flux.
The net magnetization  near  boundaries
grows monotonically with increasing $\theta$ as shown in Fig.~\ref{chern}(b), which
is equivalent to the spin transfer 
being pumped from the left edge to the right edge without accumulating in the bulk.
As shown in Fig.~\ref{chern}(c), a spin pump linearly increases with $\theta$ 
on $L_y=6$ cylinder,
which leads to a quantized net spin transfer $\Delta S^z|_{\rm edge} = 0.500$ at $\theta = 2\pi$
and a quantized Chern number $C = 1/2$,
fully characterizing the state as the $\nu = 1/2$ FQH state \cite{SR_4_6317, PRL_71_3697, PRB_52_4223}.
For the system with $L_y=4$ as shown in the inset of Fig.~\ref{chern}(c), we find some
finite size effect as the 
spin pump initially is zero for small $\theta$, which jumps to the expected values of the linear pumping behavior
at a larger $\theta$ for $J_{xy}^{\prime} = 0.1,0.2$.
The CSL is protected by the finite bulk excitation gap (shown later in Fig.~\ref{spectrum}) and grows stronger with increasing
system width. Based on the quantized Chern number established on $L_y=6$ cylinder with different geometries  \cite{suppl}
and the conformal edge spectrum for the groundstates,
we find a robust CSL phase for $J_{xy}^{\prime} \gtrsim 0.06$ as shown in the phase diagram Fig.~\ref{chern}(d) for $J_z=0$.
In the critical SL region, we  observe strong magnetization fluctuations in the bulk during the process of inserting flux in consistent
with the collapsing of the neutral excitation gap.

\begin{figure}
\includegraphics[width=0.9\linewidth]{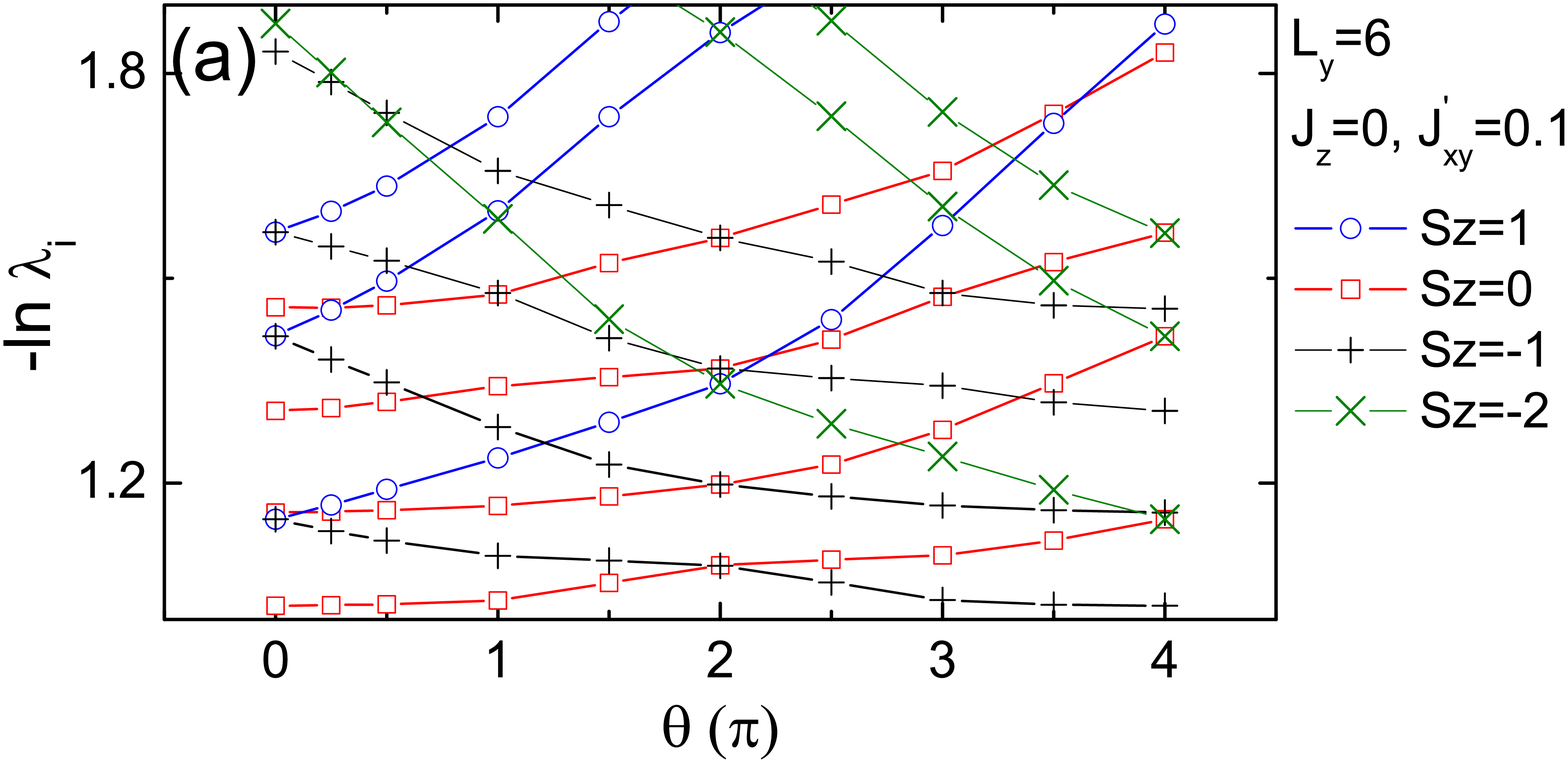}
\includegraphics[width=1.0\linewidth]{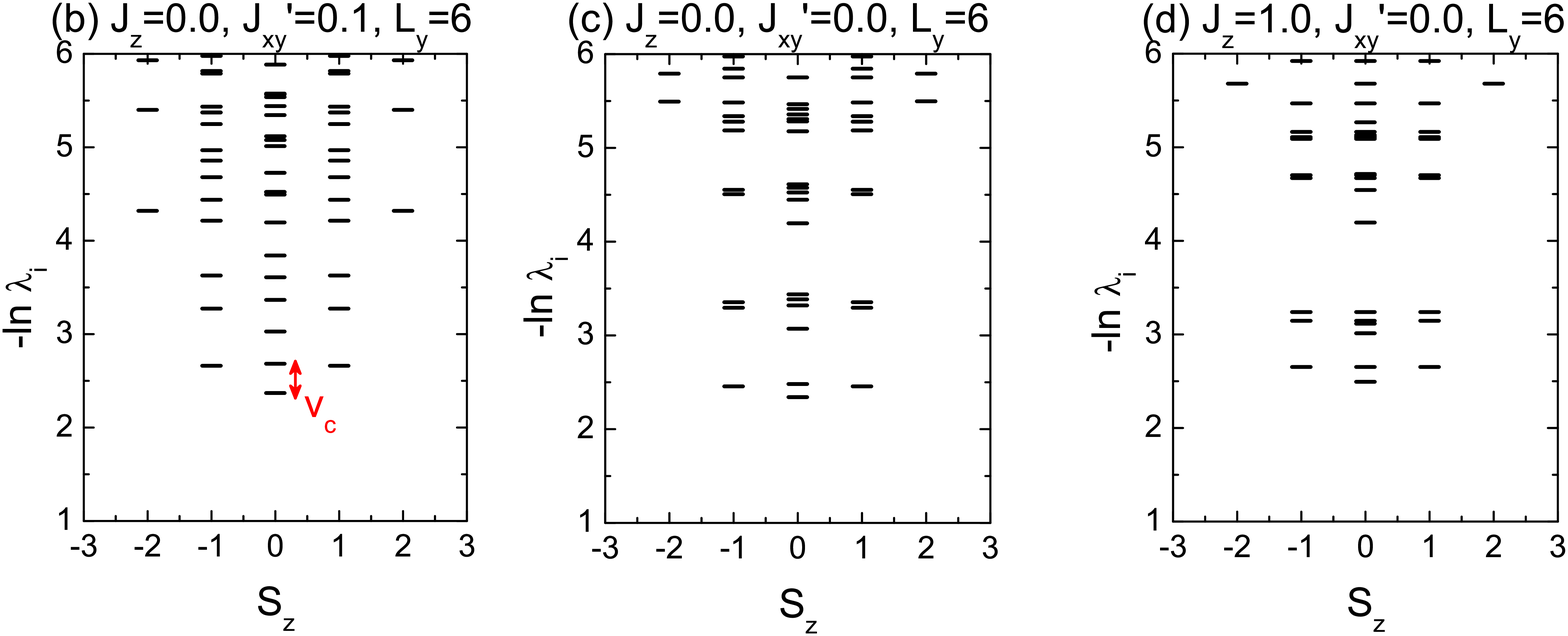}
\caption{(color online) (a) ES flow with inserting flux $\theta$ for $J_z = 0, J^{\prime}_{xy} = 0.1$ on $L_y = 6$ cylinder.
ES for (b) $J_z=0.0, J^{\prime}_{xy}=0.1$, (c) $J_z=0.0, J^{\prime}_{xy}=0.0$,
and (d) $J_z=1.0,J^{\prime}_{xy}=0.0$ on $L_y=6$ cylinder. The arrow denoted by $v_c$ in (b) indicates the two lowest eigenvalues
in the $S^z=0$ sector which are used to calculate the chiral edge-mode velocity.}\label{flow}
\end{figure}

\begin{figure}
\includegraphics[width=0.9\linewidth]{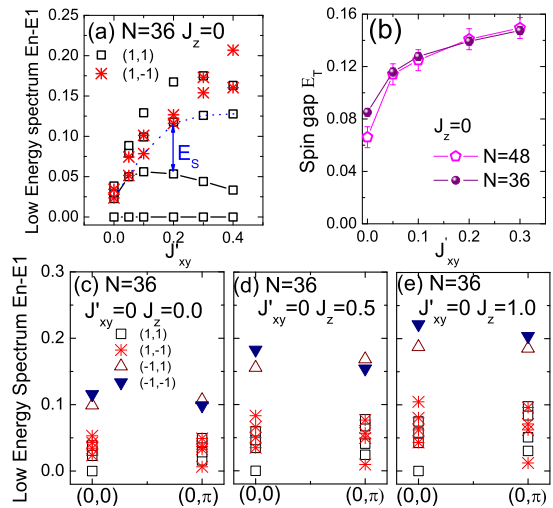}
\caption{(color online) (a) Evolution of the ED low-energy spectrum in the $k=(0,0)$ sector with $J_{xy}^{\prime}$ for
the XY model ($J_z=0$) on the $N=3\times 3\times 4$ torus. The singlet gap between the lowest two  groundstates and
higher-energy states is denoted as $E_s$. (b) $J_{xy}^{\prime}$ dependence of spin gap $E_T$ for the XY model on the $N=3\times 3\times 4$
and $3\times 4\times 4$ tori. Low-energy spectrum of the NN XXZ model ($J^{\prime}_{xy}=0$) at $k=(0,0)$ and $(0,\pi)$
sectors for (c) $J_z = 0.0$, (d) $J_z=0.5$, and (e) $J_z=1.0$ on the $3\times 3\times 4$ torus.
The label $(\pm 1,\pm 1)$ denote the quantum numbers related to spin inversion and lattice $\pi-$rotation symmetries.} \label{spectrum}
\end{figure}

\textit{Entanglement spectrum flow.---}
The  CSL and the critical   state can be understood  based on
the response of ES to inserted  flux \cite{JSM, 1407.6985}.
For a CSL at $J_{xy}^{\prime}=0.1$, as shown in Fig.~\ref{flow}(a), the eigenvalues of the reduced density matrix
are degenerating about the  $\pm S^z$ sectors  at $\theta=0$.
By increasing $\theta$, the spectrum lines in the positive $S^z$ sectors flow up  continuously,
while those in the negative $S^z$ sectors flow down (this is selected by
the sign of Chern number due to spontaneous TRS breaking).
At $\theta=2\pi$, the eigenvalues in the $S^z=0$ and $S^z=-1$ sectors become degenerate.
As a result, after inserting a flux quantum, a net spin transfer
$\Delta S|_L = \langle S^z|_L \rangle = \sum_i \lambda_i S^z_i=-1/2$ is realized and the spectrum
becomes symmetric about $S^z=-1/2$.
Thus the ES flow directly detects the gapless feature in the edge spectrum through inserting flux.
By inserting $4\pi$ flux, the ES continues to flow and, it
becomes symmetric about $S^z=-1$ at $\theta=4\pi$, indicating two spinons have been transferred from
the left edge to the right edge while the bulk of the system goes back to the vacuum sector.

With decreasing $J^{\prime}_{xy}$, the robust Chern number quantization and the spectrum flow persist to $J_{xy}^{\prime} \simeq 0.06$.
By following the evolution of the ES, we find that
the CSL is becoming less strong at smaller $J^{\prime}_{xy}$,  where
the chiral velocity (proportional to the gap between the lowest two spectrum levels in the $S^z=0$ sector as indicated in Fig.~\ref{flow}(b)) \cite{suppl}
diminishes with decreasing $J'_{xy}$.
As illustrated in Figs.~\ref{flow}(b) and \ref{flow}(c), we observe that the ES as a function of
quantum number $S^z$ before and after
the phase transition appear to be similar at $J_{xy}^{\prime}=0.1$ and $0.0$.
However, they are significantly different in momentum space.
The spectrum for $J^{\prime}_{xy} = 0.1$ preserves the same robust conformal chiral edge spectrum as demonstrated
in Fig.~\ref{Fig1}(c) with many entanglement eigenstates  carrying nonzero $k_y$ \cite{suppl}.
However, once the phase transition takes place,
the groundstate wavefunction has TRS, and the low-lying entanglement states
shown in Fig.~\ref{flow}(c) have the momentum quantum number $k_y$
either $0$ or $\pi$ if they are nondegenerate,
which comes from the mixing between eigenstates 
with opposite chiralities.
Furthermore, these low-lying eigenstates do not  respond to the inserted flux.
The mixing of entanglement states with opposite chiralities illustrates what happens to the physical edge states \cite{qi}.
These edge states with opposite chiralities also mix and merge into the bulk
and become the low energy gapless excitations in the bulk. 
These observations are consistent with
the field theory description for the quantum phase transition between two states with different Chern numbers \cite{PRL_84_3950, PRB_53_15845}.
Interestingly,  the ES for the NN kagome Heisenberg model in Fig.~\ref{flow}(d) is similar to the one of the NN XY model.

\textit{Low-energy excitations.---}
We first study the evolution of the low-energy singlet excitations in the $S^z=0$
sector as a function of $J_{xy}^{\prime}$ for
the XY model ($J_z=0$).
As shown in Fig.~\ref{spectrum}(a) of the spectrum for $36$-site torus system in $k=(0,0)$ sector, we find
two low-energy near degenerate
groundstates separated by a finite singlet gap $E_s$ from higher energy excitations
in the CSL phase at larger  $J^{\prime}_{xy} \simeq  0.3$ side \cite{SR_4_6317}.
With decreasing $J_{xy}^{\prime}$,
the singlet  gap $E_s$ reduces, which collapses to vanishing small  at $J_{xy}^{\prime} \sim 0 - 0.05$.
Thus the quantum phase transition from the CSL to the TRS preserving state is driven by such a neutral excitation gap closing.
For comparison,  we also obtain the triplet gap $E_T$ in DMRG calculations using torus systems with
$N=3\times 4\times 3$ and $3\times 4\times 4$ as shown in Fig.~\ref{spectrum}(b) \cite{DMRG}.
Similarly, $E_T$ drops with reducing $J_{xy}^{\prime}$ and it becomes much smaller
for $N=48$ system at $J'_{xy}=0$.
Thus, our results indicate that the critical state is centered near $J^{\prime}_{xy}=0$,
where the singlet gap vanishes and the spin gap is very tiny or vanished ($E_T$ reduces with $N$
and $E_T\simeq 0.049$ for $N=3\times 5\times 6$).
The appearance of low-energy singlet excitations below the finite-size spin gap can be understood as the gapless neutral mode for the topological
quantum phase transition \cite{PRL_84_3950, PRB_53_15845}, which necessarily exists for such a transition.

We further study the whole phase diagram with varying $J_z$ and $J'_{xy}$, where similar CSL to critical phase transition
is observed as illustrated in the phase diagram Fig.~\ref{Fig1}(a).
Furthermore, we examine the low energy spectra of the NN XXZ model ($J^{\prime}_{xy}=0$) on $N=3\times 4\times 3$ torus system.
As shown in Figs.~\ref{spectrum}(c)-\ref{spectrum}(e) for $J_z=0.0 ,0.5$ and $1.0$,
we find near continuous low energy excitations \cite{EPJB_2_501} collapsing together below the spin triplet gap,
which implies the gapless singlet excitations in the whole critical SL region.
The structure of the energy spectra remains very similar,
which indicates that the spin interaction $J_z$ term may only enhance the energy scale of excitations.
The gapped SL \cite{Science_332_1173} may exist neighboring with the critical SL
close to the NN Heisenberg model and  we cannot determine the precise phase boundary
of the critical SL due to the limited system width we can access in DMRG simulations.

\textit{Summary and discussions.---}
We identify a TRS broken CSL phase with a small pertubation $J^{\prime}_{xy} \simeq 0.06$ in the $J_{xy}-J'_{xy}$ XY model,
while the NN XY model is in a critical phase adjacent to the CSL with vanishing singlet excitations
and a small or vanishing spin triplet gap based on ED and DMRG studies.
Furthermore, by studying the evolution of ES crossing the quantum phase transition,
we identify that the quantum phase transition takes place through  the coupling and mixing
of the chiral  states with opposite chiralities, which
naturally lead to a critical state with TRS and gapless neutral excitations. The quantum phase transition appears
to be very smooth, which is driven by the continuous closing
of the gap for spin singlet excitations.
However, it is important to mention that limited by the range of systems one can access
using DMRG, one cannot determine
if there is a discontinuity in the singlet gap at the transition point  in thermodynamic limit.
Thus the quantum phase transition can be  a   weakly first order or continuous transition,
which demands study based on effective theory for these novel SL states.
Finally, we find that the NN $J_z$ coupling leads to a
phase diagram with an extended regime for the critical SL possibly including or close to the NN Heisenberg kagome model.
While the neutral excitation has to be gapless in such a critical SL,  we find that the spin gap is possibly
finite, but very small, which grows bigger with the increase of $J_z$.
Our DMRG calculations for spin gap on larger tori
and ED calculations for singlet gap indicate that these gaps
decrease towards the NN models with reducing $J^{\prime}_{xy}$, which are  more
consistent with a critical SL. However,
the gapped $Z_2$ SL \cite{Science_332_1173} may develop on larger systems in the NN models
through opening the vison gap from the gapless neutral excitations outside the critical SL.
This open and challenging question is desired to be addressed in the future based on different
numerical methods  and effective field theory approaches.


We hope to thank  L.~Balents, F. D. M. Haldane, L.~Fu, P. A. Lee, T.~Senthil,
Z. Y. Weng,  and especially X. G. Wen for stimulating discussions.
We also thank Y.~C. He and L.~Cincio for their insightful discussions
about developing infinite DMRG algorithm.
This research is supported by the U.S. Department of Energy, Office of Basic Energy Sciences under
grant No. DE-FG02-06ER46305 (W.Z., D.N.S.), the National Science Foundation through grants
DMR-1408560 (S.S.G). 


\clearpage

\begin{appendices}

\section{Supplemental Material}

\section{Infinite DMRG algorithm}
We also use the infinite density matrix renormalization group (iDMRG) \cite{0804.2509} to study this model.
In the iDMRG algorithm, we first start from a small system size.
Then we insert one column in the center and optimize the energy by sweeping the inserted column.
After the optimization, we absorb the new column into the original existing system and get the new boundary Hamiltonians.
We repeat the inserting, optimizing and absorbing procedure until the energy convergence is achieved.
Compared with the finite DMRG simulation, iDMRG grows the lattice by one column at each iteration and only sweeps the
inserted column, thus the computation cost is significantly reduced.
iDMRG is especially efficient to deal with the gapped topological order system, which
allows us to obtain the ground states with well-defined anyonic flux as first proposed in Ref. \cite{PRL_110_067208}.
In our work, we have confirmed that the iDMRG obtains the fixed-point ground state wavefunction in the center of
cylinder that is exactly the same as that obtained from the finite DMRG simulations (the same energy and the identical
entanglement spectrum within the numerical error.).

\section{Entanglement spectrum on XC geometry}

\begin{figure}
   \centering
   \includegraphics[width=0.7 \linewidth]{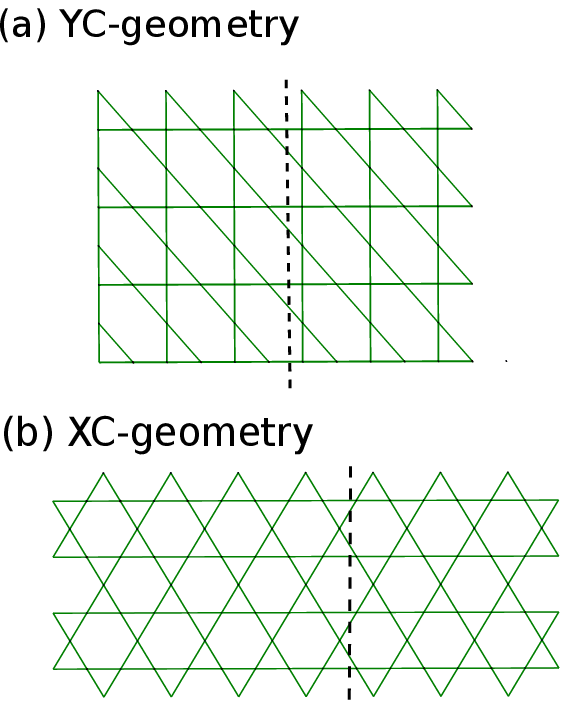}
   \caption{(color online) Kagome cylinder on (a) YC geometry and (b) XC geometry. The cylinders are closed in the $y$
direction and opened in the $x$ direction.}\label{Geo}
\end{figure}

\begin{figure}
   \centering
   \includegraphics[width=1.0 \linewidth]{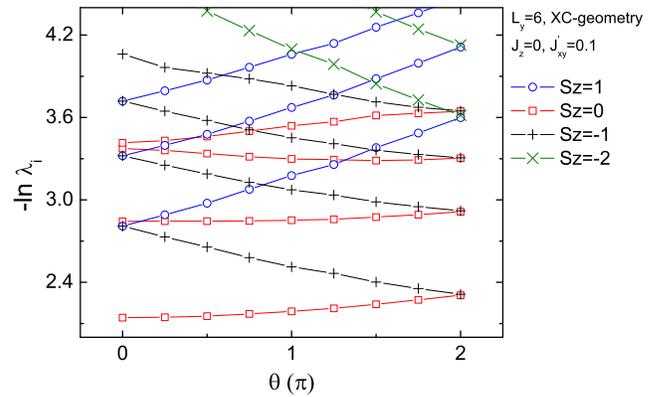}
   \caption{(color online) Entanglement spectrum flow for $J_z=0.0, J^{\prime}_{xy} = 0.1$ on $L_y = 6$ XC cylinder, which is obtained by keeping $3000$ states. }\label{flow_XC}
\end{figure}

There are two kinds of cylinder geometry on kagome lattice often being studied in DMRG, YC-geometry in SFig. \ref{Geo}(a)
and XC-geometry in SFig. \ref{Geo}(b).
In the main text, the demonstrated results are all based on YC-geometry.
Here we show that the entanglement spectrum and the spectrum flow shown on YC-geometry are
robust and insensitive to the lattice geometry.
In SFig.~\ref{flow_XC}, we demonstrate the entanglement spectrum flow for $J_z=0, J^{\prime}_{xy}=0.1$ on $L_y=6$ XC-geometry.
The features of the spectrum flow are consistent with the results on YC-geometry shown in Fig. 3(a) of the main text.
The eigenvalues in $S^z=1$ and $S^z=-1$ sectors are degenerate at $\theta=0$. By increasing flux $\theta$,
the eigenvalues in $S^z=1$ sector flow up while those in $S^z=-1$ sector flow down  continuously.
At $\theta=2\pi$, the eigenvalues in the $S^z=0$ and $S^z=-1$ sectors become degenerate, which results in a fractionally
quantized Chern number $C=1/2$.
Thus, the phase diagram shown in the main text is robust for different geometries.

\begin{figure}
   \centering
   \includegraphics[width=1.0\linewidth]{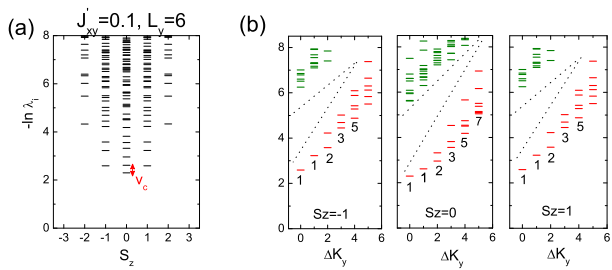}
   \caption{(color online) (a) Entanglement spectrum for $J_z=0.0, J^{\prime}_{xy} = 0.1$ on $L_y = 6$ YC cylinder,
which is obtained by keeping $3000$ states. The red arrow denoted as $v_c$ indicates the two lowest values in $k=0$ and $2\pi/L_y$ sectors that
are used to calculate the chiral velocity. (b) Near degenerating pattern for the low-lying entanglement spectra with different
relative momentum $\Delta k_y$ and total spin $S^z$ for the same system in (a). $\Delta k_y$ is in unit of $2\pi /6$.} \label{ES}
\end{figure}

\section{Chiral Velocity From Entanglement Spectrum}

Entanglement spectrum resembles the edge excitation spectrum that can be viewed as a fingerprint of topological order.
In the main text, we show the entanglement spectrum at $J_z=0.0, J^{\prime}_{xy}=0.3$ in Figs. 1(c) and 1(d). The characteristic chiral edge spectrum
indicates the chiral spin liquid (CSL) state. Here we show the spectrum at $J_z=0.0, J^{\prime}_{xy}=0.1$ on YC-geometry,
which is closer to the phase boundary $J^{\prime}_{xy} \simeq 0.06 J_{xy}$.
As shown in SFig. \ref{ES}, the entanglement spectrum also exhibits the degeneracy pattern
\{1,1,2,3,5,...\} consistent with the CSL state.

Entanglement spectrum gives not only the characteristic degeneracy pattern of the edge excitation,
but also the edge-mode velocity \cite{PRB_80_235330}.
In the conformal field theory, the edge-mode of
the Laughlin state is described by a single branch of chiral charged bosons.
The velocity of the charged bosons $v_c$ is not an universal quantity as it depends on the microscopic interaction
and edge confinement.
In the cylinder geometry, we can define $v_c$ through the lowest values of entanglement spectrum with edge momentum $k=0$ and
$k=\frac{2\pi}{L_y}$:
$v_c=\frac{\Delta E}{\hbar \Delta k}=\frac{E_0(k=2\pi/L_y)-E_0(k=0)}{\hbar 2\pi/L_y}$, where $E_0(k)=min\{-ln\lambda(k)\}$
is the lowest eigenvalue with edge momentum $k$.
Thus, from our results on YC-geometry, we have $\frac{v_c(J^{\prime}_{xy}=0.1)}{v_c(J^{\prime}_{xy}=0.3)}\approx 0.439$.
Although the CSL state is still robust at $J^{\prime}_{xy}=0.1$, the edge-mode velocity is reduced compared to $J^{\prime}_{xy}=0.3$
in the deep CSL phase. With further decreasing $J^{\prime}_{xy}$, we find that the difference of the lowest value between
the momentum sectors $k=0$ and $k=2\pi/L_y$ continuously decreases before the quantum phase transition takes place.
The reducing of chiral velocity obtained from entanglement spectrum is related with the drop of the bulk excitation gap
with decreasing $J'_{xy}$, which is consistent with a very weakly first order transition or a continuous transition
driven by the collapsing of the bulk gap and the destruction
of the gapless edge states at the same time.


\end{appendices}


\begin{thebibliography}{99}

\bibitem{Nature_464_199} L. Balents,
Nature \textbf{464}, 199 (2010).


\bibitem{PRB_40_7387} X. G. Wen,
Phys. Rev. B \textbf{40}, 7387 (1989).

\bibitem{PRB_41_9377} X. G. Wen and Q. Niu,
Phys. Rev. B \textbf{41}, 9377 (1990).

\bibitem{IJMPB_4_239} X. G. Wen,
Int. J. Mod. Phys. B \textbf{4}, 239 (1990).

\bibitem{PRB_82_155138} X. Chen, Z. C. Gu, and X. G. Wen,
Phys. Rev. B \textbf{82}, 155138 (2010).

\bibitem{Science_235_1196} P. W. Anderson,
Science \textbf{235}, 1196 (1987).

\bibitem{PRL_61_2376} D. S. Rokhsar and S. A. Kivelson,
Phys. Rev. Lett. \textbf{61}, 2376 (1988).

\bibitem{PRL_66_1773} N. Read and S. Sachdev,
Phys. Rev. Lett. \textbf{66}, 1773 (1991).

\bibitem{PRB_44_2664} X. G. Wen,
Phys. Rev. B \textbf{44}, 2664 (1991).

\bibitem{PRB_62_7850} T. Senthil and M. P. A. Fisher,
Phys. Rev. B \textbf{62}, 7850 (2000); Phys. Rev. Lett. \textbf{86}, 292 (2001).

\bibitem{PRL_86_1881} R. Moessner and S. L. Sondhi,
Phys. Rev. Lett. \textbf{86}, 1881 (2001).

\bibitem{PRB_65_224412} L. Balents, M. P. A. Fisher, and S. M. Girvin,
Phys. Rev. B \textbf{65}, 224412 (2002).

\bibitem{PRB_66_205104} T. Senthil and O. I. Motrunich,
Phys. Rev. B \textbf{66}, 205104 (2002).

\bibitem{PRL_89_277004} O. I. Motrunich and T. Senthil,
Phys. Rev. Lett. \textbf{89}, 277004 (2002).

\bibitem{PRL_94_146805} D. N. Sheng and L. Balents,
Phys. Rev. Lett. \textbf{94}, 146805 (2005).

\bibitem{RMP_78_17} P. A. Lee, N. Nagaosa, and X. G. Wen,
Rev. Mod. Phys. \textbf{78}, 17 (2006).

\bibitem{AP_321_2} A. Kitaev,
Ann. Phys. (N.Y.) \textbf{321}, 2 (2006).

\bibitem{PRL_99_097202} D. F. Schroeter, E. Kapit, R. Thomale, and M. Greiter,
Phys. Rev. Lett. \textbf{99}, 097202 (2007).

\bibitem{Science_321_1306} P. A. Lee,
Science \textbf{321}, 1306 (2008).

\bibitem{NP_7_772} S. V. Isakov, M. B. Hastings, R. G. Melko,
Nat. Phy. \textbf{7}, 772 (2011).

\bibitem{PRL_108_247206} H. Yao and S. A. Kivelson,
Phys. Rev. Lett. \textbf{108}, 247206 (2012).

\bibitem{PRB_82_024419} F. Wang,
Phys. Rev. B \textbf{82}, 024419 (2010).

\bibitem{PRB_84_024420} Y. M. Lu and Y. Ran,
Phys. Rev. B \textbf{84}, 024420 (2011).

\bibitem{PRL_107_087204} B. K. Clark, D. A. Abanin, and S. L. Sondhi,
Phys. Rev. Lett. \textbf{107}, 087204 (2011).

\bibitem{PRB_86_024424} H. C. Jiang, H. Yao, and L. Balents,
Phys. Rev. B \textbf{86}, 024424 (2012).

\bibitem{PRL_110_127203} R. Ganesh, J. V. D. Brink, and S. Nishimoto,
Phys. Rev. Lett. \textbf{110}, 127203 (2013).

\bibitem{PRL_110_127205} Z. Y. Zhu, D. A. Huse, and S. R. White,
Phys. Rev. Lett. \textbf{110}, 127205 (2013).

\bibitem{PRB_88_165138} S. S. Gong, D. N. Sheng, O. I. Motrunich, and M. P. A. Fisher,
Phys. Rev. B \textbf{88}, 165138 (2013).

\bibitem{PRL_113_027201} S. S. Gong, W. Zhu, D. N. Sheng, O. I. Motrunich, and M. P. A. Fisher,
Phys. Rev. Lett. \textbf{113}, 027201 (2014).


\bibitem{PRL_101_117203} H. C. Jiang, Z. Y. Weng, and D. N. Sheng,
Phys. Rev. Lett. \textbf{101}, 117203 (2008).

\bibitem{Science_332_1173} S. Yan, D. Huse, and S. R. White,
Science \textbf{332}, 1173 (2011).

\bibitem{PRL_109_067201} S. Depenbrock, I. P. McCulloch, and U. Schollw\"{o}ck,
Phys. Rev. Lett. \textbf{109}, 067201 (2012).

\bibitem{NP_8_902} H. C. Jiang, Z. H. Wang, and L. Balents,
Nat. Phy. \textbf{8}, 902 (2012).

\bibitem{PRL_96_110404} A. Kitaev, J. Preskill, Phys. Rev. Lett. \textbf{96}, 110404 (2006).

\bibitem{PRL_96_110405} M. Levin, X. G. Wen, Phys. Rev. Lett. \textbf{96}, 110405 (2006).

\bibitem{PRB_60_1654} L. Balents, M. P. A. Fisher, and C. Nayak,
Phys. Rev. B \textbf{60}, 1654 (1999).

\bibitem{PRB_89_075110} Y. C. He, D. N. Sheng, and Y. Chen,
Phys. Rev. B \textbf{89}, 075110 (2014).

\bibitem{PRL_98_117205} Y. Ran, M. Hermele, P. A. Lee, and X. G. Wen,
Phys. Rev. Lett. \textbf{98}, 117205 (2007).

\bibitem{PRB_84_020407} Y. Iqbal, F. Becca, and D. Poilblanc,
Phys. Rev. B \textbf{84}, 020407(R) (2011).

\bibitem{SR_4_6317} S. S. Gong, W. Zhu, and D. N. Sheng,
Sci. Rep. \textbf{4}, 6317 (2014).

\bibitem{PRL_112_137202} Y. C. He, D. N. Sheng, and Y. Chen,
Phys. Rev. Lett. \textbf{112}, 137202 (2014).

\bibitem{PRL_50_1395} R. B. Laughlin,
Phys. Rev. Lett. \textbf{50}, 1395 (1983).

\bibitem{PRL_59_2095} V. Kalmeyer and R. B. Laughlin,
Phys. Rev. Lett. \textbf{59}, 2095 (1987).

\bibitem{PRB_39_11413} X. G. Wen, F. Wilczek, and A. Zee,
Phys. Rev. B \textbf{39}, 11413 (1989).

\bibitem{PRL_70_2641} K. Yang, L. K. Warman, and S. M. Girvin,
Phys. Rev. Lett. \textbf{70}, 2641 (1993).

\bibitem{PRB_52_4223} F. D. M. Haldane and D. P. Arovas,
Phys. Rev. B \textbf{52}, 4223 (1995).

\bibitem{1407.0869} J. W. Mei and X. G. Wen,
arXiv:1407.0869.

\bibitem{1407.2740} Y. C. He and Y. Chen,
arXiv:1407.2740.

\bibitem{1401.3017} B. Bauer, L. Cincio, B. P. Keller, M. Dolfi, G. Vidal, S. Trebst, A. W. W. Ludwig,
arXiv:1401.3017.

\bibitem{1307.8194} M. Barkeshli,
arXiv:1307.8194.

\bibitem{PRL_91_2003} Y. Shimizu, K. Miyagawa, K. Kanoda, M. Maesato, and G. Saito,
Phys. Rev. Lett. \textbf{91}, 107001 (2003).

\bibitem{PRL_95_177001} Y. Kurosaki, Y. Shimizu, K. Miyagawa, K. Kanoda, and G. Saito,
Phys. Rev. Lett. \textbf{95}, 177001 (2005).

\bibitem{PRB_77_2008} T. Itou, A. Oyamada, S. Maegawa, M. Tamura, and R. Kato,
Phys. Rev. B \textbf{77}, 104413 (2008).

\bibitem{PRL_98_077204} P. Mendels, F. Bert, M. A. de Vries, A. Olariu, A. Harrison, F. Duc, J. C. Trombe, J. S. Lord, A. Amato, and C. Baines,
Phys. Rev. Lett. \textbf{98}, 077204 (2007).

\bibitem{PRL_98_107204} J. S. Helton, K. Matan, M. P. Shores, E. A. Nytko, B. M. Bartlett, Y. Yoshida, Y. Takano, A. Suslov, Y. Qiu, J.-H. Chung,
D. G. Nocera, and Y. S. Lee,
Phys. Rev. Lett. \textbf{98}, 107204 (2007).

\bibitem{PRL_103_237201} M. A. de Vries, J. R. Stewart, P. P. Deen, J. O. Piatek, G. J. Nilsen, H. M. Rønnow, and A. Harrison,
Phys. Rev. Lett. \textbf{103}, 237201 (2009).

\bibitem{PRL_109_037208} B. F\r{a}k, E. Kermarrec, L. Messio, B. Bernu, C. Lhuillier, F. Bert, P. Mendels,
B. Koteswararao, F. Bouquet, J. Ollivier, A. D. Hillier, A. Amato, R. H. Colman, and A. S. Wills,
Phys. Rev. Lett. \textbf{109}, 037208 (2012).

\bibitem{Nature_492_7429} T. H. Han, J. S. Helton, S. Chu, D. G. Nocera, J. A. Rodriguez-Rivera, C. Broholm, and Y. S. Lee,
Nature \textbf{492}, 7429 (2012).

\bibitem{PRL_110_207208} L. Clark, J. C. Orain, F. Bert, M. A. De Vries, F. H. Aidoudi, R. E. Morris, P. Lightfoot, J. S. Lord, M. T. F. Telling, P. Bonville, J. P. Attfield, P. Mendels, and A. Harrison,
Phys. Rev. Lett. \textbf{110}, 207208 (2013).

\bibitem{PRB_53_15845} I. A. McDonald and F. D. M. Haldane,
Phys. Rev. B \textbf{53}, 15845 (1996).

\bibitem{PRL_84_3950} X. G. Wen,
Phys. Rev. Lett. \textbf{84}, 3950 (2000).

\bibitem{PRL_69_2863} S. R. White,
Phys. Rev. Lett. \textbf{69}, 2863 (1992).

\bibitem{suppl} See more information in Supplemental Material.

\bibitem{Vidal} L. Cincio and G. Vidal,
Phys. Rev. Lett. \textbf{110}, 067208 (2013).

\bibitem{PRL_110_236801} Michael P. Zaletel, Roger S. K. Mong, and Frank Pollmann,
Phys. Rev. Lett. \textbf{110}, 236801 (2013).

\bibitem{CFT_book} P. Di Francesco, P. Mathieu, and D. Senechal,
\textit{Conformal Field Theory} (Springer, New York, 1997), Chap. 15.6.

\bibitem{PRL_101_010504} H. Li and F. D. M. Haldane,
Phys. Rev. Lett. \textbf{101}, 010504 (2008).

\bibitem{PRL_71_3697} Y. Hatsugai,
Phys. Rev. Lett. \textbf{71}, 3697 (1993).

\bibitem{JSM} W. Zhu, S. S. Gong, and D. N. Sheng,
J. Stat. Mech. 2014 (8), P08012.

\bibitem{1407.6985} A. G. Grushin, J. Motruk, M. P. Zaletel, and F. Pollmann,
arXiv:1407.6985.

\bibitem{qi} X. L.  Qi, H. Katsura, A. W. W. Ludwig,
Phys. Rev. Lett. \textbf{108}, 196402 (2012).

\bibitem{DMRG} While one cannot get a few low energy singlet excitations in DMRG accurately,  one can obtain the
triplet gap $E_T$ by targeting the ground states in the different total $S^z$ sectors separately.

\bibitem{EPJB_2_501} C. Waldtmann, H. U. Everts, B. Bernu, C. Lhuillier, P. Sindzingre, P. Lecheminant, L. Pierre,
Eur. Phys. J. B \textbf{2}, 501 (1998).


\end{thebibliography}

\begin{thebibliography}{99}

\bibitem{0804.2509} I. P. McCulloch, arXiv:0804.2509.

\bibitem{PRL_110_067208} L. Cincio and G. Vidal,
Phys. Rev. Lett. \textbf{110}, 067208 (2013).

\bibitem{PRB_80_235330}Z. X. Hu, E. H. Rezayi, X. Wan, and K. Yang, Phys. Rev. B {\bf 80}, 235330 (2009).

\end{thebibliography}
\end{document}